\documentclass[preprint,showpacs,preprintnumbers,amsmath,amssymb]{revtex4-1}

\usepackage{graphicx}
\usepackage{dcolumn}
\usepackage{bm}

\begin{document}

\title{Fabrication and characterization of an induced GaAs single hole transistor}

\author{O. Klochan}
\email{klochan@phys.unsw.edu.au}
\author{J.C.H. Chen}
\author{A.P. Micolich}
\author{A.R. Hamilton}
\email{Alex.Hamilton@unsw.edu.au}
 \affiliation{School of Physics, University of New South Wales,
 Sydney NSW 2052, Australia.}

\author{K. Muraki}
\affiliation{NTT Basic Research Laboratory, NTT Corporation, 3-1
Morinosato Wakamiya, Atsugi, Kanagawa 243-0198, Japan}

\author{Y. Hirayama}
\affiliation{Department of Physics, Tohoku University, 6-3 Aramaki
aza Aoba, Aobaku Sendai, Miyagi 980-8578, Japan}

\date{\today}

\begin{abstract}
We have fabricated and characterized a single hole transistor in an
undoped AlGaAs-GaAs heterostructure. Our device consists of a p-type
quantum dot, populated using an electric field rather than
modulation doping. Low temperature transport measurements reveal
periodic conductance oscillations due to Coulomb blockade. We find
that the low frequency charge noise is comparable to that in
modulation-doped GaAs single electron transistors (SETs), and an
order of magnitude better than in silicon SETs.
\end{abstract}
\maketitle

The ability to isolate and control single spins in solid state
systems is a problem of significant interest. Although systems such
as P nuclei in Si \cite{KaneNature98} and Nitrogen-vacancy centers
in diamond \cite{WrachtrupJPCM06} have emerged, electron quantum
dots defined using surface-gate techniques on AlGaAs/GaAs
heterostructures have, due to their purity and maturity, been the
focus of most attention.~\cite{ElzermanNature04, KoppensNature06,
KouwenhovenPSS06, HansonRMP07} However, a major limitation of
electron quantum dots is randomization of the electron's spin
orientation due to hyperfine interactions with Ga and As nuclei.
There are two material-based ways to overcome this problem. One is
to use materials with spin zero nuclei such as $^{12}$C or
$^{28}$Si. The alternative is to retain the advantages of the
AlGaAs/GaAs system by using holes rather than electrons. Holes
experience a much weaker hyperfine coupling, resulting in a
significantly enhanced spin coherence lifetime,~\cite{BulaevPRL07}
as demonstrated by optical measurements of self-assembled hole
quantum dots.~\cite{BrunnerScience09} However, reports on the
electronic properties of hole quantum dots are
scarce,~\cite{GrbicAPL05, KomijaniEPL08} due to the poor stability
of surface-gate defined hole devices.~\cite{EnsslinNatPhys06,
Taskinen09}

We have previously shown that it is possible to make stable hole
quantum wires using an `induced' approach where the device is
populated electrostatically using a biased gate rather than by
modulation doping.~\cite{ClarkeJAP06, KlochanAPL06} In this letter,
we extend this technique to realize a stable hole quantum dot, which
we demonstrate by operating as single hole transistor. This type of
device could in the future be used as a system for isolating,
manipulating and studying single hole spins.

Our devices were fabricated on a (311)A-oriented AlGaAs/GaAs
heterostructure featuring (from undoped GaAs buffer upwards): a
$175$~nm undoped AlGaAs barrier, a $25$~nm GaAs spacer, and a
$75$~nm p$^{+}$-GaAs cap (degenerately doped with Si as an acceptor
to $5\times10^{18}$ cm$^{-3}$) used as a metallic gate. Annealed
AuBe Ohmic contacts, fabricated using a self-aligned
process,~\cite{KaneAPL93} provide electrical contact to the
two-dimensional hole system (2DHS) that forms at the AlGaAs/GaAs
interface. The p$^{+}$-gate layer is divided into seven independent
gates using electron-beam lithography and shallow wet etching, as
shown Fig.~1(a) inset. The largest gate is the top gate, which leads
to an accumulation of holes when negatively biased to $V_{TG}
\lesssim -0.1$~V, populating the dot and the adjacent 2DHS
reservoirs. The four gates in the corners define two quantum point
contacts (denoted L and R) used to control the transparency of the
leads connecting the quantum dot to the reservoirs. The two plunger
gates (PG) are used to control the number of holes in the dot. The
resulting dot has lithographic dimensions $790 \times 720$~nm.
Electrical studies were performed in a dilution refrigerator, with a
minimum hole temperature of $\sim 140$~mK. The two-terminal
conductance was measured using lock-in techniques with a $20~\mu$V
ac excitation at $5$~Hz. The series resistance was dominated by the
Ohmic contacts, which had a resistance of $\sim 40$ k$\Omega$.

\begin{figure}[h]
\includegraphics[width = 8.5cm]{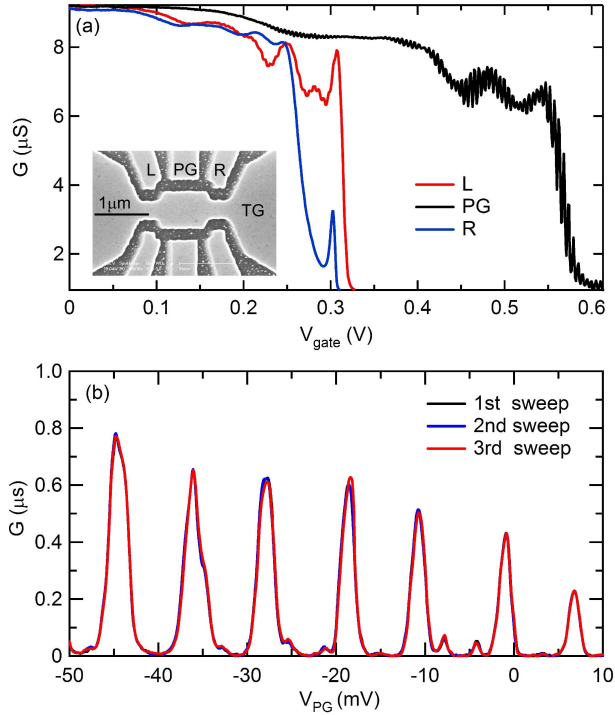}
\caption{\label{fig1} (a) Two terminal conductance $G$ versus left
QPC (red), right QPC (blue) and plunger gate (black) bias with
$V_{TG}=-0.3$~V and magnetic field $B = 0$~T. The inset shows a
scanning electron micrograph of the device, etched regions are black
and the seven gates are grey. (b) Three consecutive measurements of
$G$ versus plunger gate bias $V_{PG}$ at $V_{TG} = -0.38$~V, $V_{L}
= 0.58$~V and $V_{R} = 0.62$~V demonstrating device stability. The
data in (b) was obtained at $B = 0.5$~T.}
\end{figure}

We begin in Fig.~1(a) by characterizing the quantum point contact
(QPC) and plunger gates. The top-gate bias was set to $-0.3$~V to
give a hole density $p = 9.2\times 10^{10}$~cm$^{-2}$ in the 2D
reservoirs and populate the dot. At this density, the mobility in
the reservoirs was $590,000$~cm$^{2}/$Vs. Increasing the positive
bias on the left or right QPCs generates a series of steps in the
conductance $G$ as the 1D subbands in the dot's entrance and exit
leads are depopulated. The pinch-off voltages (i.e., where $G$
becomes zero) for the two QPCs are identical to within $10\%$.
Increasing $V_{PG}$ also causes steps in conductance, and leads to
Coulomb Blockade (CB) oscillations with a period of $\sim 5$~mV.
These CB oscillations persist even with the QPC gates unbiased, in
contrast to modulation-doped devices. The oscillations become more
prominent at more positive bias because capacitive coupling between
the plunger gate and the entrance and exit constrictions makes them
less transparent.

To enhance the CB oscillations we increase $V_{TG}$ to $-0.38$~V ($p
= 1.26 \times 10^{11}$~cm$^{-2}$) and then increase $V_{L}$ and
$V_{R}$ until the entrance and exit leads are fully closed and no
current passes through the dot. A small negative bias is then
applied to the bottom plunger gate (top plunger gate remains
unbiased), which has two effects: it opens the entrance and exit
leads slightly, and tunes the hole density (i.e., occupancy) of the
dot. In Fig.~1(b) we plot $G$ versus bottom plunger gate bias, and
obtain well-defined CB peaks separated by regions of zero
conductance. To demonstrate the stability of our device, we show
sweeps of $V_{PG}$ repeated for a second (blue) and third (red)
time, which show no change in the peak locations in $V_{PG}$ and
almost negligible variations in peak amplitude. For $V_{PG} <
-50$~mV (i.e., beyond the left axis in Fig.~1(a)) the peak amplitude
increases rapidly and the conductance minima rise above $G = 0$,
similar to the behavior at $V_{PG} \approx 0.6$~V in Fig.~1(a). We
observe a variation in the period of oscillations from $7.6$ to
$9.9$~mV, similar to that in Ref.~\cite{GrbicAPL05}. In addition to
the primary CB oscillations we also observe higher frequency
oscillations with reduced amplitude. In order to confirm that the
smaller higher frequency peaks are not resonances caused by
background impurities, we applied a small perpendicular field $B =
0.5$~T; the peaks were unaffected by this. It is also unlikely that
these peaks arise because the dot has broken into two or more
smaller dots, since the smaller dots would have a smaller
capacitance to the gates and hence show lower frequency
oscillations, contrary to what we observe. One possibility is that
the two sets of CB oscillations are due to the interplay between the
light-hole and heavy-hole energy levels in the entrance and exit
QPCs: in our 2D hole system the ground state is heavy-hole-like,
whereas in a tightly confined 1D system the ground state is
light-hole-like.~\cite{CsontosPRB07} This will be the focus of a
more extended study in the future.

\begin{figure}[h]
\includegraphics[width = 8.5cm]{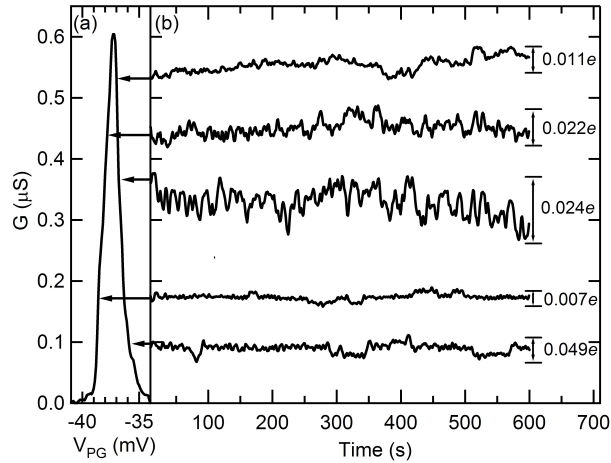}
\caption{\label{fig2} (a) Conductance $G$ vs (a) plunger gate bias
$V_{PG}$; and (b) $G$ vs time $t$ for several set values of $V_{PG}$
(the corresponding locations on the CB peak in (a) are indicated by
the arrows). In both cases $V_{TG} = -0.38$~V and $B = 0.5$~T. The
vertical scale bars in (b) show the maximum magnitude of the
peak-to-peak charge fluctuations for each time trace.}
\end{figure}

A common characteristic of modulation-doped hole quantum devices is
poor temporal stability, with a tendency for the conductance to
drift over long time scales even when the gate biases are all held
constant.~\cite{DaneshvarPRB97} This is particularly undesirable for
quantum dot devices, which are often used as highly sensitive
electrometers due to their high
transconductance.~\cite{FujisawaScience06} The conductance
fluctuations are typically due to charge noise caused by ionized
dopants, which we seek to minimize with the undoped device design
presented here. To further gauge the effectiveness of our approach,
in Fig.~2 we present stability tests performed by setting five
different operating points $V_{PG}$ (Fig.~2(a)) on the CB peak
centered at $V_{PG} = -36$ mV in Fig. 1(b), and measuring $G$ over a
ten minute period (see Fig.~2(b)). In each case, the average
conductance remains approximately constant, despite some random
fluctuations. These fluctuations are small, corresponding to a
charge fluctuation between $0.007$ and $0.05$ of an electron over
the ten minute period. We extract the power spectral density
$S_{q}(f)$ from an analysis of these fluctuations, obtaining values
ranging from $2$ to $8 \times 10^{-4}$~$e/\sqrt{Hz}$ at a frequency
$f = 0.1$~Hz. This is almost an order of magnitude lower than the
$S_{q} \approx 1.5\times 10^{-3}$~$e/\sqrt{Hz}$ at $0.1$~Hz measured
in undoped Si SETs,~\cite{HourdakisAPL08} and compares very
favorably with the $S_{q} = 2\times 10^{-4}$~$e/\sqrt{Hz}$ at $3$~Hz
measured in modulation-doped GaAs radio-frequency
SETs.~\cite{FujisawaAPL01} Note that $1/f$-noise becomes more
pronounced at lower frequencies, such that if equal $S_{q}$ values
are obtained in two separate devices $1$ and $2$ at different
frequencies $f_{1} << f_{2}$, then device $1$ is significantly
quieter. This highlights the superior noise performance that can be
obtained with our induced GaAs single-hole transistor technology.

It is interesting to note evidence of random telegraph signals in
the bottom two traces in Fig.~2(b). In modulation-doped devices this
has been attributed to ionization/deionization of dopants in the
modulation-doping layer or current leakage from the surface
gates.~\cite{BuizertPRL08} These mechanisms should be greatly
reduced in our device due to the absence of modulation doping and
the deliberate use of small top-gate bias to minimize the gate
leakage current (measured as $< 2$~pA at $V_{TG} = -0.38$~V).
Surface-state traps in the vicinity of the $75$~nm deep trench that
was etched to define the gates are one possible origin of this
noise. However, these traps would be at least $160$~nm from the dot,
and their contribution is thus likely to be small. An alternative is
that background impurities may have ionized during the
low-temperature illumination of the device to improve the Ohmic
contact resistance. Further studies utilizing higher bandwidth
measurement techniques,~\cite{TaskinenRSI08} and surface
passivation~\cite{BessolovSemi98} may help in determining the origin
of the remaining noise in our devices.

\begin{figure}[h]
\includegraphics[width = 8.5cm]{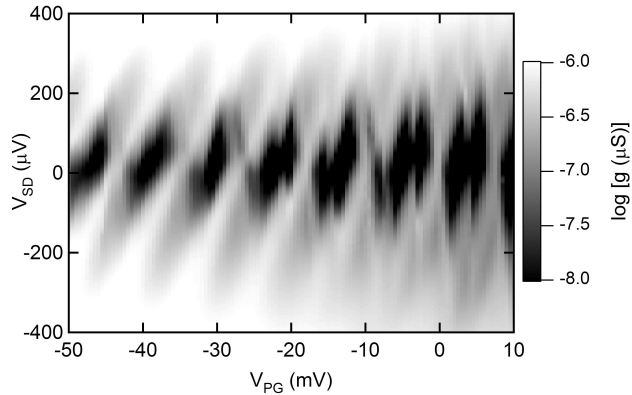}
\caption{\label{fig3} Grayscale plot of the  differential
conductance $g$ (color axis) versus plunger gate bias $V_{PG}$
($x$-axis) and applied dc source-drain bias $V_{SD}$ ($y$-axis),
obtained with $V_{TG} = -0.38$~V and $B = 0.5$~T. The dark/light
regions correspond to low/high conductance. Coulomb diamonds are
clearly observed and appear tilted due to asymmetrically applied
source-drain bias.}
\end{figure}

As a final characterization, we have used source-drain bias
spectroscopy to determine the key energy scales for our dot. In
Fig.~3 we present a grayscale plot of the differential conductance
$g = dI/dV_{SD}$ versus $V_{PG}$ and applied dc source-drain bias
$V_{SD}$, with Coulomb diamonds clearly evident. We note the
presence of additional diamond-like features, which are related to
the smaller high-frequency peaks in Fig.~1(b) rather than charge
noise. We extract a charging energy $E_{C}$ ranging from $150$ to
$200$~$\mu$eV from an analysis of the Coulomb diamonds,
corresponding to a total dot capacitance of $8 - 10$~aF. This is
more than double that expected from the sum of the individual gate
capacitances ($4.2$~aF), suggesting significant capacitive coupling
between the dot and the adjacent 2D reservoirs. From the dot's
lithographic dimensions and hole density, we estimate the dot
occupancy to be less than $460$ holes, giving a single particle
energy spacing $\Delta E \approx 4~\mu$eV. This is substantially
smaller than $k_{B}T \sim 12~\mu$eV, explaining why we do not
observe excited state transport in our device (cf.
Ref.~\cite{KomijaniEPL08}). Reducing the size of our dots to allow
lower occupancy operation and the observation of excited-state
transport, whilst maintaining their low noise characteristics, will
be the focus of future work.

In summary, we have reported the fabrication of a GaAs single hole
transistor using a heterostructure without modulation-doping. Our
device shows stable electrical characteristics with little drift in
conductance at fixed gate bias and significantly improved noise
performance compared to both modulation-doped GaAs and undoped Si
single electron transistors. The dot design we present here holds
much promise for future studies of single hole spin manipulation.

This work was funded by the Australian Research Council (ARC). ARH
was supported by an ARC Professorial Fellowship. We thank T. P.
Martin and L. H. Ho for helpful discussions and J. Cochrane for
technical support.

\end{document}